\documentstyle[aps,prl,twocolumn,epsf]{revtex}
\begin{document}
\draft
\wideabs{
\title{Phase-fluctuating 3D condensates in elongated traps}
\author{D.S. Petrov${}^{1,3}$, G.V. Shlyapnikov${}^{1,2,3}$, and J.T.M.
Walraven${}^{1}$}
 
\address{${}^1$ FOM Institute for Atomic and Molecular
Physics, Kruislaan 407,   1098 SJ Amsterdam, The Netherlands \\ 
${}^2$ Laboratoire Kastler Brossel*, Ecole Normale Superieure, 24 rue Lhomond,
F-75231 Paris Cedex 05, France \\ 
${}^3$ Russian Research Center, Kurchatov
Institute,   Kurchatov Square, 123182 Moscow, Russia}

\date{\today}
\maketitle
\begin{abstract}
We find that in very elongated 3D trapped Bose gases, even at temperatures far
below the BEC transition temperature $T_c$, the equilibrium state will be a 3D
condensate with fluctuating phase (quasicondensate). At sufficiently low
temperatures the phase fluctuations are suppressed and the quasicondensate
turns into a true condensate. The presence of the phase fluctuations allows
for extending thermometry of Bose-condensed gases well below those established
in current experiments.
\end{abstract}
 }
  
\footnotetext{ 
\footnotemark  LKB is an unit\'{e} de recherche de l'Ecole Normale 
Sup\'{e}rieure et de l'Universit\'{e} Pierre et Marie Curie, associ\'{e}e 
au CNRS. 
} 
\narrowtext
 
Phase coherence properties are among the most interesting aspects of 
Bose-condensed gases. Since the discovery of Bose-Einstein condensation 
(BEC) in trapped ultra-cold clouds of alkali atoms \cite{discovery}, 
various experiments have proved the presence of phase coherence in trapped 
condensates. The MIT group \cite{Ket} has found the interference of two 
independently prepared condensates, once they expand and overlap after 
switching off the traps.  The MIT
\cite{KetBragg}, NIST \cite{Phi} and Munich \cite{Ess} experiments provide 
evidence for the phase coherence of trapped condensates through the
measurement of the phase coherence length and/or single particle
correlations.  

These results support the usual picture of BEC in 3D gases. In equilibrium,
the fluctuations of density and phase are important only in a narrow 
temperature range near the BEC transition temperature $T_c$.
Outside this region, the fluctuations are suppressed and the condensate 
is phase coherent. This picture precludes the interesting
physics of phase-fluctuating condensates, which is present in 2D and  1D
systems (see \cite{petrov2,petrov1} and refs. therein). 

In this Letter we show that the phase coherence properties of 3D Thomas-Fermi
(TF) condensates depend on their shape. In very elongated 3D condensates, the
axial phase fluctuations are found to manifest themselves even at temperatures
far below $T_c$. Then, as the density fluctuations are suppressed,
the equilibrium state will be a {\it condensate with fluctuating phase}
(quasicondensate) similar to that in 1D trapped gases \cite{petrov1}.
Decreasing $T$ below a sufficiently low temperature, the 3D
quasicondensate gradually turns into a true condensate. 

The presence and the temperature dependence of axial phase fluctuations
in sufficiently elongated 3D condensates suggests a principle of thermometry 
for Bose-condensed gases with indiscernible thermal clouds. The idea is to
extract the temperature from a measurement of the axial phase fluctuations,
for example by measuring the single-particle correlation function. This
principle works for quasicondensates or for any condensate that can
be elongated adiabatically until the phase fluctuations become observable.

So far, axial phase fluctuations have not been measured in experiments with 
cigar-shaped condensates. We discuss the current experimental situation and 
suggest how one should select the parameters of the
cloud in order to observe the phase-fluctuating 3D condensates. 

We first consider a 3D Bose gas in an elongated cylindrical harmonic
trap and analyze the behavior of the single-particle correlation function.
The natural assumption of the existence of a true
condensate at $T=0$ automatically comes out of these calculations. 
In the TF regime, where the mean-field (repulsive) interparticle
interaction greatly exceeds the radial ($\omega_{\rho}$) and axial
($\omega_z$) trap frequencies, the density profile of the zero-temperature
condensate has the well-known shape $n_0(\rho,z)=n_{0m}
(1-\rho^2/R^2-z^2/L^2)$, where $n_{0m}=\mu/g$ is the maximum
condensate density, with $\mu$ being the chemical potential,
$g=4\pi\hbar^2a/m$, $m$ the atom mass, and $a>0$ the scattering length.
Under the condition $\omega_{\rho}\gg\omega_z$, the radial size of the
condensate, $R=(2\mu/m\omega_{\rho}^2)^{1/2}$, is much smaller  than the axial
size $L=(2\mu/m\omega_z^2)^{1/2}$.   

Fluctuations of the density and phase of the condensate,
in particular at finite $T$, are related to elementary excitations of
the cloud. The density fluctuations are dominated by the excitations with 
energies of the order of $\mu$. The wavelength of these excitations is much
smaller than the radial size of the condensate. Hence, the density
fluctuations have the ordinary 3D character and are small. Therefore, one can
write the total field operator of atoms as $\hat\psi({\bf r})=\sqrt{n_0({\bf
r})}\exp(i\hat\phi({\bf r}))$, where $\hat\phi({\bf r})$ is the operator of
the phase. The single-particle correlation function is then
expressed through the mean square fluctuations of the phase (see, e.g.
\cite{Popov}):    \begin{equation}  \label{opdm} 
\!\langle\hat\psi^{\dagger}({\bf r})\hat\psi({\bf
r}')\rangle\!=\!\sqrt{n_0({\bf r})n_0({\bf r}')} 
\exp\{-\langle[\delta\hat\phi({\bf r},{\bf r}')]^2\rangle/2\},\!
\end{equation} 
with $\delta\hat\phi({\bf r},{\bf r}')=\hat\phi({\bf r})-\hat\phi({\bf
r}')$. The operator $\hat\phi({\bf r})$ is given by (see, e.g., \cite{Shev})
\begin{equation}      \label{operphi}    
\hat\phi({\bf r})=[4n_0({\bf r})]^{-1/2}\sum_{\nu}f_\nu^{+}({\bf r})\hat a_\nu 
+h.c.,
\end{equation}
where $ \hat{a}_\nu$ is the annihilation operator of the excitation with
quantum number(s) $\nu$ and energy $\epsilon_\nu$, $f_\nu^{+}= u_\nu +
v_\nu$, and the $u,v$ functions of the excitations are determined by the
Bogolyubov-de Gennes equations.

The excitations of elongated condensates can be divided into two groups:
``low energy'' axial excitations with energies
$\epsilon_{\nu}<\hbar\omega_{\rho}$, and ``high energy'' excitations with
$\epsilon_{\nu}>\hbar\omega_\rho$. The latter have 3D
character as their wavelengths are smaller than the radial size $R$. 
Therefore, as in ordinary 3D condensates, these excitations
can only provide small phase fluctuations. The ``low-energy'' axial excitations
have wavelengths larger than $R$ and exhibit a pronounced 1D behavior. Hence,
one expects that these excitations give the most important contribution to the
long-wave axial fluctuations of the phase. 

The solution of the Bogolyubov-de Gennes equations for the low-energy axial 
modes gives the spectrum $\epsilon_j=\hbar\omega_z\sqrt{j(j+3)/4}$ 
\cite{stringari}, where $j$ is a positive integer. The wavefunctions
$f_j^+$ of these modes have the form
\begin{equation} 
\label{fpm} 
f_j^{+}({\bf r})=\sqrt{\frac{(j+2)(2j+3)gn_0({\bf r})}{4\pi
(j+1)R^2L\epsilon_j}}P_j^{(1,1)}\left(\frac{z}{L}\right),  
\end{equation}
where $P_j^{(1,1)}$ are Jacobi polynomials. Note that the
contribution of the low-energy axial excitations to the phase operator
(\ref{operphi}) is independent of the radial coordinate $\rho$. 

Relying on Eqs.~(\ref{operphi}) and (\ref{fpm}), we now calculate the mean
square axial fluctuations of the phase at distances $|z-z'|\ll R$. 
As in 1D trapped gases \cite{petrov1}, the vacuum fluctuations are
small for any realistic axial size $L$. The thermal fluctuations are
determined by the equation  
\begin{eqnarray} \label{sqfl}
\lefteqn{\langle[\delta\hat\phi(z,z')]^2\rangle_T=
\sum_{j=1}^{\infty}\frac{\pi\mu (j+2)(2j+3)}{15(j+1)\epsilon_j
N_0}\times}\nonumber\hspace{8cm}\\
\left(P_j^{(1,1)}\left(\frac{z}{L}\right)-P_j^{(1,1)}
\left(\frac{z'}{L}\right)\right)^2 N_j,   
\end{eqnarray}
with $N_0=(8\pi/15)n_{0m}R^2L$ being the number of Bose-condensed
particles, and $N_j$ the equilibrium occupation numbers
for the excitations. The main contribution to the sum over $j$ in
Eq.(\ref{sqfl}) comes from several lowest excitation modes, and at 
temperatures $T\gg\hbar\omega_z$ we may put $N_j=T/\epsilon_j$. Then, 
in the central part of the cloud ($|z|,|z'|\ll L$) a straightforward
calculation yields    
\begin{equation}   \label{center}
\langle[\delta\hat\phi(z,z')]^2\rangle_T=\delta_L^2|z-z'|/L,
\end{equation}
where the quantity $\delta_L^2$ represents the phase fluctuations on a distance
scale $|z-z'|\sim L$ and is given by
\begin{equation}    \label{deltaL}
\delta_L^2(T)=32\mu T/15N_0(\hbar\omega_z)^2.
\end{equation}
Note that at any $z$ and $z'$ the ratio of the phase correlator (\ref{sqfl})
to  $\delta_L^2$ is a universal function of $z/L$ and $z'/L$: 
\begin{equation}     \label{f}
\langle[\delta\hat\phi(z,z')]^2\rangle_T=\delta_L^2(T)f(z/L,z'/L). 
\end{equation}
In Fig.1 we present the function $f(z/L)\equiv f(z/L,-z/L)$
calculated numerically from Eq.(\ref{sqfl}).

\begin{figure} 
\epsfxsize=\hsize 
\epsfbox{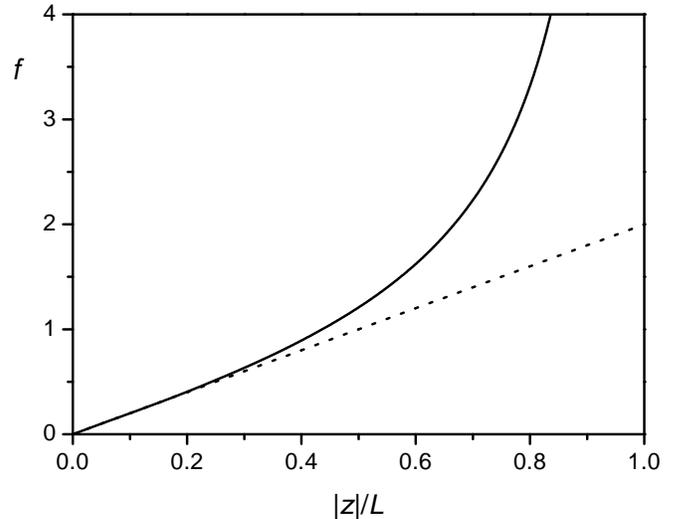} 
\caption{\protect
The function $f(z/L)$. 
The solid curve shows the result of our numerical calculation, and the dotted
line is $f(z)=2|z|/L$ following from Eq.(\ref{center}).}  
\label{1}  \end{figure} 

The phase fluctuations decrease with temperature. As the TF chemical potential
is $\mu\!=\!(15N_0g/\!\pi\!)^{2/5}(m{\bar \omega}^2\!/8)^{3/5}$
(${\bar \omega}=\omega_{\rho}^{2/3}\omega_z^{1/3}$), Eq.(\ref{deltaL}) can be
rewritten in the form  
\begin{equation}             \label{deltaL1}
\delta_L^2=(T/T_c)(N/N_0)^{3/5}\delta_c^2,
\end{equation}
where $T_c\approx N^{1/3}\hbar{\bar \omega}$ is the BEC transition
temperature, and $N$ the total number of particles. The presence of the 3D BEC 
transition in elongated traps requires the inequality $T_c\gg\hbar\omega_{\rho}$ 
and, hence, limits the aspect ratio to $\omega_{\rho}/\omega_z\ll N$. 
The parameter $\delta_c^2$ is given by  
\begin{equation}    \label{TT}  
\delta_c^2=\frac{32\mu(N_0=N)}{15N^{2/3}\hbar{\bar
\omega}}\left(\frac{\omega_{\rho}}{\omega_z}\right)^{4/3}\!\!\propto
\frac{a^{2/5}m^{1/5}\omega_{\rho}^{22/15}}{N^{4/15}\omega_z^{19/15}}.
\end{equation}
Except for a narrow interval of temperatures just below $T_c$, the fraction of 
non-condensed atoms is small and Eq.(\ref{deltaL1}) reduces to
$\delta_L^2=(T/T_c)\delta_c^2$. Thus, the phase fluctuations 
can be important at large values of the parameter $\delta_c^2$, whereas for
$\delta_c^2\ll 1$ they are small on any distance scale and
one has a true Bose-Einstein condensate. In Fig.2 we present the
quantity $\delta_c^2$ for the parameters of various experiments with
elongated condensates. 

In the Konstanz \cite{Rempe} and Hannover \cite{Ertmer} experiments the
ratio $T/T_c$ was smaller than $0.5$. 
In the recent experiment
\cite{Ertmer1} the value $\delta_c^2\approx 3$ has been reached, but the
temperature was very low. Hence, the axial phase fluctuations were rather
small in these experiments and they were dealing with true condensates. The
last statement also holds for the ENS experiment \cite{Dalibard} where $T$ was
close to $T_c$ and the Bose-condensed fraction was $N_0/N\approx 0.75$.
 
The single-particle correlation function is determined by Eq.(\ref{opdm})
only if the condensate density $n_0$ is much larger than the density of
non-condensed atoms, $n'$. Otherwise, this equation should be completed by
terms describing correlations in the thermal cloud. However, 
irrespective of the relation between $n_0$ and $n'$, 
Eq.(\ref{opdm}) and Eqs.~(\ref{sqfl})-(\ref{deltaL}) correctly
describe phase correlations in the condensate as long as the fluctuations of
the condensate density are suppressed. This is still the case for $T$ 
close to $T_c$ and $N_0\ll N$, if we do not enter the region of
critical fluctuations. Then, Eq.(\ref{deltaL1}) gives
$\delta_L^2=(N/N_0)^{3/5}$. At the highest temperatures of the Bose-condensed
cloud in the MIT sodium experiment \cite{Ketterle}, the condensed fraction was
$N_0/N\sim 0.1$ and the phase fluctuations were still small. 

\begin{figure} 
\epsfxsize=\hsize 
\epsfbox{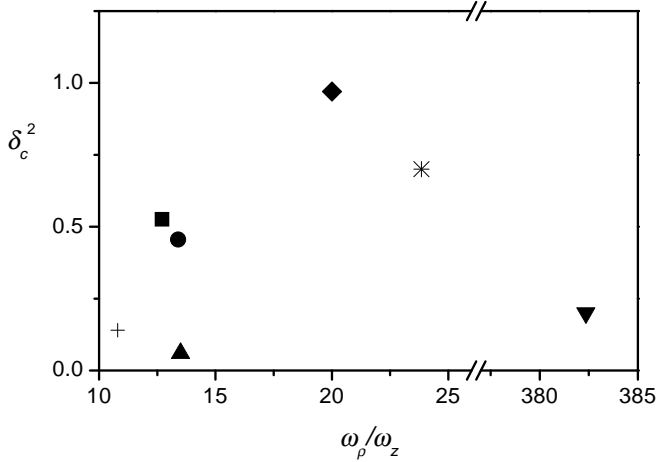}  
\caption{\protect
The parameter $\delta_c^2$ for experiments with elongated condensates. 
The up and down triangles stand for $\delta_c^2$ in the sodium
[11] and hydrogen [12] MIT experiments, respectively.
Square, cross,  diamond, circle, and star show $\delta_c^2$ for the rubidium
experiments at  Konstanz [13], Munich [5], Hannover [14],
ENS [15], and AMOLF [16].}
\label{2}   
\end{figure}

On the contrary, for $N_0/N\approx 0.06$ in the hydrogen experiment
\cite{Kleppner}, with $\delta_c^2$ from Fig.2 we estimate $\delta_L^2\approx
1$. The same or even larger value of $\delta_L^2$ was reached in
the Munich Rb experiment \cite{Ess} where the gas temperature was varying in a
wide interval around $T_c$. In the Rb experiment at AMOLF \cite{Walraven}, the
smallest observed Bose-condensed fraction was $N_0/N\approx 0.03$, which
corresponds to $\delta_L^2\approx 5$. However, axial fluctuations of the
phase have not been measured in these experiments. 

We will focus our attention on the case
where $N_0\approx N$ and the presence of the axial phase fluctuations is
governed by the parameter $\delta_c^2$.   
For $\delta_c^2\gg 1$, the nature of the Bose-condensed state depends on
temperature. In this case we can introduce a characteristic temperature
\begin{equation} \label{Tph}
T_{\phi}=15(\hbar\omega_z)^2N/32\mu
\end{equation}   
at which the quantity $\delta_L^2\approx 1$ (for $N_0\approx N$).
In the temperature interval $T_{\phi}<T<T_c$, the phase fluctuates on a
distance scale smaller than $L$. Thus, as the density fluctuations are
suppressed, the Bose-condensed state is a condensate with fluctuating
phase or quasicondensate. The expression for the radius of phase
fluctuations (phase coherence length) follows from Eq.(\ref{center}) and is 
given by 
\begin{equation}     \label{rphi}
l_{\phi}\approx L(T_{\phi}/T).  
\end{equation}
The phase coherence length $l_{\phi}$ greatly exceeds the correlation length
$l_c=\hbar/\sqrt{m\mu}$. Eqs.~(\ref{rphi}) and (\ref{Tph}) give the ratio
$l_{\phi}/l_c\approx (T_c/T)(T_c/\hbar\omega_{\rho})^2\gg 1$. Therefore, the
quasicondensate has the same density profile and local correlation properties
as the true condensate.  However, the phase coherence properties of
quasicondensates will be drastically different (see below).

The decrease of temperature to well below $T_{\phi}$ makes the phase
fluctuations small ($\delta_L^2\ll 1$) and continuously transforms 
the quasicondensate into a true condensate. 

It is interesting to compare the described behavior of the interacting gas 
for $\delta_c^2\gg 1$, with the two-step BEC predicted for the ideal Bose gas 
in elongated traps \cite{KvD}. In both cases, at $T_c$ the particles 
Bose-condense in the ground state of the radial motion. However, the ideal 
gas remains non-condensed (thermal) in the axial direction for
$T>T_{1D}=N\hbar\omega_z/\ln{2N}$ (assuming $T_{1D}<T_c$), and there is a
sharp cross-over to the axial BEC regime at $T\approx T_{1D}$. The interacting
Bose gas below $T_c$ forms the 3D TF (non-fluctuating) density profile, and
the spatial correlations become non-classical in all directions. For
$\delta_c^2\gg 1$, the axial phase fluctuations at $T\sim T_c$ are still
large, and one has a quasicondensate which continuously transforms into a true
condensate at $T$ below $T_{\phi}$. Note that $T_{\phi}$ is quite different
from $T_{1D}$ of the ideal gas.  

Let us now demonstrate that 3D elongated quasicondensates can be achieved for
realistic parameters of trapped gases. As found above, the existence of a
quasicondensate requires large values of the parameter $\delta_c^2$ given by
Eq.(\ref{TT}). Most important is the dependence of $\delta_c^2$ on the aspect 
ratio of the cloud $\omega_{\rho}/\omega_z$, whereas the dependence on the 
number of atoms and on the scattering length is comparatively weak. Fig.3 
shows $T_c/T_{\phi}=\delta_c^2$, $\mu/T_{\phi}$, and the temperature
$T_{\phi}$ as functions of $\omega_{\rho}/\omega_z$ for
rubidium condensates at $N=10^5$ and $\omega_{\rho}=500$ Hz.
Comparing the results for $\delta_c^2$ in Fig.3 with the data in Fig.2, we see
that 3D quasicondensates can be obtained by transforming the presently
achieved BEC's to more elongated geometries corresponding to 
$\omega_{\rho}/\omega_z\agt 50$.

One can distinguish between quasicondensates and true BEC's in various types
of experiments. By using the Bragg spectroscopy method developed at MIT
one can measure the momentum distribution of particles in the trapped gas
and extract the coherence length $l_{\phi}$\cite{KetBragg}. The use of two
(axially) counter-propagating laser beams to absorb a photon from one beam
and emit it into the other one, results in axial momentum transfer to the
atoms which have momenta at Doppler shifted resonance with the beams.
These atoms form a small cloud which will axially separate from the rest 
of the sample provided the mean free path greatly exceeds the axial size $L$.
The latter condition can be assured by applying the Bragg excitation after
abruptly switching off the radial confinement of the trap. The axial momentum
distribution is then conserved if the dynamic evolution of the cloud does not
induce axial velocities. According to the scaling approach
\cite{Kagan}, this is the case for the axial frequency decreasing as
$\omega_z(t)=\omega_z(0)[1+\omega_{\rho}^2t^2]^{-1/2}$.

\begin{figure} 
\epsfxsize=\hsize 
\epsfbox{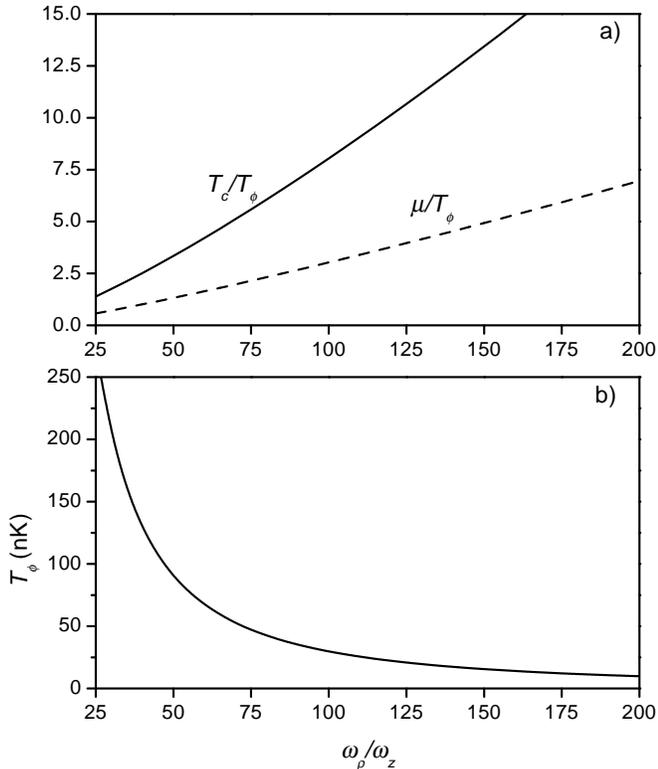} 
\caption{\protect
The ratios $T_c/T_{\phi}=\delta_c^2$ and $\mu/T_{\phi}$
in (a) and the temperature $T_{\phi}$ in (b), versus the aspect ratio
$\omega_{\rho}/\omega_z$ for trapped Rb condensates with
$N=10^5$ and $\omega_{\rho} =500$ Hz.}   
\label{3}   \end{figure}

In "juggling" experiments described in \cite{petrov1} and similar
to those at NIST and Munich \cite{Phi,Ess}, one can directly measure the 
single-particle correlation function. The latter is obtained by repeatedly 
ejecting small clouds of atoms from the parts $z$ and $z'$ of the sample 
and averaging the pattern of interference between them in the detection
region over a large set of measurements. As follows from Eqs.~(\ref{deltaL})
and (\ref{f}), for $z'=-z$ the correlation function depends on temperature as
$\exp{\{-\delta_L^2(T)f(z/L)/2\}}$, where $f(z/L)$ is given in Fig.1.

The phase fluctuations are very sensitive to temperature. From
Fig.3 we see that one can have $T_{\phi}/T_c<0.1$,
and the phase fluctuations are still significant at $T<\mu$, where only a 
tiny indiscernible thermal cloud is present. 

This suggests a principle for thermometry of 3D Bose-condensed gases with 
indiscernible thermal clouds. If the sample is not an elongated quasicondensate
by itself, it is first transformed to this state by adiabatically increasing
the aspect ratio $\omega_{\rho}/\omega_z$. This does not change the ratio
$T/T_c$ as long as the condensate remains in the 3D TF regime. Second, the
phase coherence length $l_{\phi}$ or the single-particle correlation function
are measured. These quantities depend on temperature if the latter is of the
order of $T_{\phi}$ or larger. One thus can measure the ratio $T/T_c$ for the
initial cloud,  which is as small as the ratio $T_{\phi}/T_c$ for the
elongated cloud. 

We believe that the studies of phase coherence in elongated condensates will
reveal many new interesting phenomena. The measurement of phase correlators 
will allow one to study the evolution of phase coherence in the course of the
formation of a condensate out of a non-equilibrium thermal cloud. 

When completing the paper we were informed that 3D
quasicondensates were observed in Hannover \cite{Klaus}.

We acknowledge fruitful discussions with W.D. Phillips, M. Lewenstein and K.
Sengstock. This work was supported by the Nederlandse Organisatie voor 
Wetenschappelijk Onderzoek (NWO), by the Stichting voor 
Fundamenteel Onderzoek der Materie (FOM), and by  the Russian Foundation for 
Basic Research.

\end{document}